\documentclass[prb,groupedaddress,showpacs,twocolumn]{revtex4}

\usepackage{graphicx,epsfig}

\begin{document}

\title{Mott-Superfluid transition in  bosonic ladders}

\author{P. Donohue}
\email{donohue@lps.u-psud.fr} \affiliation{Laboratoire de Physique
des Solides, CNRS-UMR 85002, Universit\'e Paris--Sud, B\^at. 510,
91405 Orsay, France}

\author{T. Giamarchi}
\email{giam@lps.u-psud.fr} \affiliation{Laboratoire de Physique
des Solides, CNRS-UMR 85002, Universit\'e Paris--Sud, B\^at. 510,
91405 Orsay, France}

\date{\today}

\begin{abstract}
We show that in a commensurate bosonic ladder, a quantum phase
transition occurs between a Mott insulator and a superfluid when
interchain hopping increases. We analyse the
properties of this transition as well as the physical properties
of the two phases. We discuss the physical consequences for
experimental systems such as Josephson Junction arrays.
\end{abstract}
\pacs{}
\maketitle

The Mott transition is one of the most striking consequences of
interactions in a quantum system. Although mostly studied in fermionic
systems, it is even more spectacular in interacting bosonic
ones where the transition occurs between a Mott
insulating state where bosons are well localized on the lattice
and a superfluid state, with zero resistivity. The existence
of such a Mott transition for
bosons has been proven in one dimension
\cite{haldane_bosons} using bosonization and studied in
higher dimensions using a scaling theory \cite{fisher_boson_loc}.
Since then the physical properties of this transition have been intensely
investigated \cite{sondhi_qcp}. For this problem, one dimension is a specially interesting
case since: (i) interactions play an extremely important role leading, even for bosons, to a Luttinger liquid state;
(ii) no true superfluid condensate exists. Nevertheless a true Mott-superfluid
transition takes place whose
critical and physical properties are now well understood both at commensurate densities
and close to commensurability \cite{haldane_bosons,fisher_boson_loc,giamarchi_mott_shortrev}.
The existence and characteristics of this transition have been investigated by
various analytical and numerical techniques
\cite{batrouni_bosons_numerique,krauth_bosons_gutzwiller,elstner_pade_bosons,kuhner_bosehubbard}
and experimentally in  Josephson Junction (JJ)
arrays \cite{vanoudenaarden_josephson_mott,fazio_josephson_junction_review}.

Even if the properties of a single bosonic chain are now rather well understood,
a crucial question is how the characteristics of the transition
evolve when going from one to two dimensions.
Indeed if one considers coupled bosonic chains there is clearly a
competition between the interactions (thus the Mott phase) and the
interchain hopping that tends to delocalize the system and push it to the
superfluid phase. One way to tackle these questions analytically
is to focus on a small number of coupled chains, i.e.
on ladders. Such ladder systems have also experimental realizations in JJ arrays.
A similar approach was followed in fermionic systems for
which related questions arose
\cite{donohue_confinement_spinless_refs,lehur_ladder_crossover,tsuchiizu_confinement_spinful}.
For commensurate fermionic ladders, quite surprisingly, only a strong crossover
is found but the system remains insulating regardless of the strength
of the interchain hopping. This prompts for the question of the existence
of the Mott-superfluid transition in the bosonic ladder.
So far the physical properties of bosonic ladders have been studied only
for the incommensurate ladder
\cite{orignac_2chain_bosonic} or for a commensurability of one
boson every two sites \cite{orignac_2spinchains} for which the
ladder is essentially equivalent to an anisotropic spin ladder \cite{schulz_spins,strong_spinchains_long}.
However the case of one boson per site, where the commensurability being
higher leads to a stronger and more involved competition between the Mott phase and the interchain
hopping is still to be understood.

We thus investigate in the present paper a bosonic ladder, at
or close to the commensurate filling of one boson per site.
We show that for this system there is a true transition,
when interchain hopping is increased, between a Mott insulator
and a superfluid phase. Large coupling expansions show that this transition
is in the Beresinskii-Kosterlitz-Thouless (BKT) universality class at
commensurate filling and in the commensurate-incommensurate one for small
doping. Various correlation functions such as the single particle one, show
universal power law decay at the transition. The transport properties,
that we compute can provide a check of the existence of this transition in
experimental systems such as the JJ arrays.

We start with the Hamiltonian
\begin{eqnarray}
H  &=&  -t \sum_{i,\alpha} (b^{\dagger}_{i,\alpha} b_{i+1,\alpha}
+ \text{h.c.}) + \frac{U}{2} \sum_{i,\alpha} n_{i,\alpha}
(n_{i,\alpha}-1) \nonumber \\
& & -\mu \sum_{i,\alpha} n_{i\alpha} -
t_\perp \sum_{i} (b^{\dagger}_{i,1} b_{i,2} + \text{h.c.})
\label{teve}
\end{eqnarray}
where $\alpha=1,2$ is the chain index. $t$ and $t_\perp$ are
respectively the intra and interchain hopping, $U$ is the on-site
particle interaction, and $\mu$ is the chemical potential. In the
following, unless we specify otherwise, $\mu$ is chosen to impose
one boson per site.   To describe the low energy properties of
(\ref{teve}) it is convenient to use the density-phase
representation of the bosons \cite{haldane_bosons}:
\begin{eqnarray}
\rho(x)\approx\rho_0-\frac{1}{\pi}\partial_x\phi+\rho_0\cos(2\phi)\\
\frac{b_i}{\sqrt{\alpha}}=\Psi(x) \propto e^{i\theta(x)}\sqrt{\rho(x)}\\
\left[\frac{-1}{\pi}\partial_x\phi(y),\theta(y')\right]=i\delta(y'-y),
\end{eqnarray}
where $\alpha$ is a short distance cut-off of the order of the
lattice spacing. Using these variables the single chain
Hamiltonian can be rewritten as:
\begin{eqnarray}
H &=& H^0 -\frac{g_u}{(2\pi\alpha)^2}\int dx \cos{2\phi} \label{eq:total} \\
H^0(u,K) &=& \frac{u}{2\pi} \int dx
\left[K(\partial_x \theta)^2+\frac{1}{K}(\partial_x \phi)^2
\right] \label{eq:quadratic}
\end{eqnarray}
All interaction effects are hidden in the Luttinger liquid
parameters $u$, the sound velocity, and $K$ a coefficient that
controls the asymptotic decay of the correlation functions. Free
bosons correspond to $K=\infty$ and hard core ones ($U=\infty$) to
$K=1$. Note that for one boson per site, the $U=\infty$ case is a
trivial band insulator. The $\cos(2\phi)$ term where $g_u\propto
U$ describes the umklapp scattering of the bosons on the lattice
and is responsible for the Mott transition
\cite{giamarchi_mott_shortrev}. Note that the continuous form
(\ref{eq:total}) is much more general than the microscopic
hamilonian (\ref{teve}) and describes systems with longer range
interactions as well. Using the symmetric basis
$\phi_{s,a}=(\phi_1\pm\phi_2)/\sqrt2$, (\ref{teve}) gives:
\begin{eqnarray}\label{eq:bosonizedhamiltonian}
H&=&H^0_a(u,K_a)+H^0_s(u,K_s)\nonumber\\
&&-\frac{2g_u}{(2\pi\alpha)^2} \int dx
\cos(\sqrt2\phi_s)\cos(\sqrt2\phi_a)\nonumber\\
&&-\frac{t_\perp}{\pi\alpha} \int dx
\cos(\sqrt2\theta_a)
\end{eqnarray}
where $H^0_{s,a}$ are defined by (\ref{eq:quadratic}). For (\ref{teve}) one has
$u_{s,a}=u$ and $K_{s,a}=K$.

\begin{figure}
\includegraphics[width=8cm]{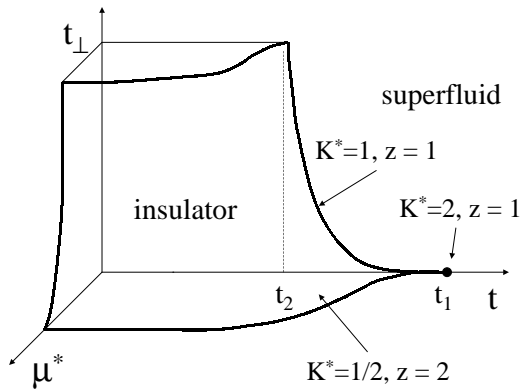}
\caption{\label{fig:phasediag} Phase diagram as a function of
$t$, $t_\perp$ and $\mu^*$ at fixed $U$. $\mu^*$ is defined as the
distance in chemical potential from commensurability (i.e. one boson per site),
that is $\mu$ minus a function of $t$ and $t_\perp$. This unimportant
shift of $\mu$ is simply to keep straight axis in the figure.
$t_1$ is the
critical value for a single chain. $t_2$ is the critical value
for an effective chain of 2 bosons/site and an interaction $U/2$
(see text). The wall of the critical surface correspond to a
commensurate-incommensurate transition with $z=2$. The phase
diagram as a function of $K$ instead of $t$ presents identical
features since for $U$ fixed, $K$ and $t$ are continuously related.}
\end{figure}

In (\ref{eq:bosonizedhamiltonian}) there is a competition between the umklapp
term, that favors localization of the charge hence order in the field $\phi_a$
(and $\phi_s$) and the interchain coupling that wants to order the relative
superfluid phase $\theta_a$ between the two chains. Since these two fields are conjugate
one can naively expect a transition between the two types of order for the antisymmetric
field. The resulting competition
affects the symmetric modes, inducing a metal-insulator
transition. To investigate this transition we use a renormalization group (RG) method
in powers of the umklapp and interchain hopping. The RG equations read:
\begin{eqnarray}\label{eq:rg}
\frac{d\tilde{g}_u}{dl} &=&(2-K_a/2-K_s/2)\tilde{g}_{u} \nonumber\\
\frac{d\tilde{t}_\perp}{dl}&=& (2-K^{-1}_a/2)\tilde{t}_\perp \nonumber \\
\frac{dK_a}{dl} &=& -\frac{K_a^2\tilde{g}_u^2}{16\pi^2}+\frac{\tilde{t}_\perp^2}{8\pi^2} \\
\frac{dK_s}{dl} &=& -\frac{K_s^2\tilde{g}_u^2}{16\pi^2} \nonumber
\end{eqnarray}
where $\tilde{g}_u=g_u/u$ and
$\tilde{t}_\perp=(4\pi^2\alpha)t_\perp/u$ are dimensionless
coupling constants. For $1/4<K<2$, (\ref{eq:rg}) show that both
$t_\perp$ and $g_u$ are relevant operators that thus tend to order
$\theta_a$ and $\phi_a$ respectively. Two different behaviors
occur depending on which operator becomes of order unity first: (i) if
the umklapp scattering eventually dominates the flow it leads to
ordering in $\phi_s$ and $\phi_a$ and we recover an insulating
behavior similar to the one of a single chain; (ii) if the interchain hopping
dominates then $\theta_a$ orders and the umklapp scattering
becomes irrelevant. The only term that can localize the symetric
component is now a term generated to second order of the form
$\cos(\sqrt8\phi_s)$. This operator has a dimension $2K$ instead
of $K$ for a single chain umklapp, and is thus much less relevant.
The precise interaction for which this operator is relevant
depends on its amplitude that can only be computed reliably when
$g_u \ll 1$. Even if this is not the case, one
can show in the large $t_\perp$ expansion below that there is a range of
interactions between $U^c_{\text{single chain}}$ and $U=\infty$
for which this operator is indeed irrelevant. Thus at fixed
interactions a transition between the Mott insulator and a
superfluid for a given value of the interchain hopping occurs.
Since both the umklapp and the interchain hopping are relevant
operators there is no weak coupling fixed point at the transition,
and the RG (\ref{eq:rg}) cannot be used to determine the
critical properties of this transition. One can however get
qualitatively the transition line as the position where the most
relevant operator changes. Using (\ref{eq:rg}) leads to
\begin{eqnarray}
\tilde{t}_{\perp}^c=(\tilde{g}_u)^{\frac{2-1/(2K)}{2-K}}
\end{eqnarray}
The resulting phase diagram is shown on Fig.~\ref{fig:phasediag}.

We now turn to the critical properties
of the insulator-superfluid transition itself. Since it is difficult to
extract from the RG flow, we can only analyse it in specific limits.
The first one is the large transverse hopping $t_\perp\gg t,U$.
On each rung there are two one-particle
states, bonding or anti-bonding. The many bosons low lying states
correspond to putting every boson in the bonding state.
The chemical potential must be such as to
ensure a particle-hole symmetry (around two bosons per ``site'')
in order to have a constant
density superfluid-insulator transition. This gives back the
familiar problem of one species of bosons on a lattice with a
commensurate filling of two bosons per site, with an effective
hopping $t$ and an on site repulsive energy $U/2$. Since the
effective interaction is reduced compared to the single chain
the large $t_\perp$ system can be superfluid even if the single
chain is Mott insulating. The Mott-superfluid transition is a
BKT one with an effective LL parameter $K_{\text{eff}}=2$ and
a dynamical exponent $z=1$
\cite{haldane_bosons,fisher_boson_loc,giamarchi_mott_shortrev}.
To obtain the
LL parameter $K_s$ of the original ladder one can compute the
correlation functions. The one-particle
green's function has a universal power law decay at the transition
$\langle\psi_{\text{eff}}^\dagger(r)\psi_{\text{eff}}(0)\rangle\propto(1/r)^{(1/4)}$.
Using (\ref{eq:correl}) leads to to the universal value $K^*_s=1$
at the transition. Since for fermionic ladders the large $t_\perp$
can lead to different phases than the small $t_\perp$ limit,
it is important to check that an identical
critical behavior is recovered in another
strong coupling limit. We rewrite (\ref{eq:bosonizedhamiltonian})
as the bosonized form of a different
microscopic lattice Hamiltonian than (\ref{teve}),
namely two coupled spin one-half
chains:
\begin{eqnarray} \label{eq:spinchains}
H &=&
H^{XXZ}_\alpha+H^{XXZ}_\beta-J^z_{\perp}\sum_iS_{\alpha,i}^zS_{\beta,i}^z
\nonumber \\
&&+h_1^x\sum_i(-1)^iS^x_{\alpha,i} \\
H^{XXZ}&=&\sum_i J (S_i^xS_{i+1}^x+S_i^yS_{i+1}^y) +J^z S_{i}^zS_{i+1}^z
\end{eqnarray}
Using the standard bozonized expressions for the spins \cite{schulz_houches_revue} one easily
recovers (\ref{eq:bosonizedhamiltonian}) from (\ref{eq:spinchains}) with
$h_x\propto t_\perp$, $J^z_\perp\propto g_u$.
$J^z_\alpha$, $J^z_\beta$ are choosen such as to recover the correct bare
Luttinger parameters $K_s$ and $K_a$.
The Hamiltonian (\ref{eq:spinchains}) can be studied in the strong coupling
limit $h_x \sim J^z_\perp \gg J_{\alpha,\beta}$ ($t_\perp \sim U \gg t$).
Each rung corresponds to a four level system that may
be easily diagonalized. Keeping only the two degenerate lowest
levels we obtain a pseudo-spin one-half chain.
If we denote its spin by $\tilde{S}$ and its couplings by $\tilde{J}$
the correspondance between
the spin operators is:
\begin{eqnarray} \label{eq:mapping}
S_\alpha^z&=&\cos(\alpha)\tilde{S}^z\\
S_\beta^z&=&\tilde{S}^z\\
S_\alpha^+ &=& \frac{1}{2}\sin(\alpha)\tilde{I}\\
S_\beta^+ &=& (-1)^{i+1}\sin(\alpha)\tilde{S}^+\\
\cos(\alpha)&=&\frac{J_\perp^z}{[(J_\perp^z)^2+(2h_1^x)^2]^{\frac{1}{2}}},
\end{eqnarray}
where $I$ is the identity.
Using these relations, (\ref{eq:spinchains}) becomes an XXZ Hamiltonian with
the effective couplings:
\begin{equation}
\frac{-\tilde{J}^z}{\tilde{J}}=\frac{J_2^z}{J_2}
+\left(\frac{J_\perp^z}{2h_1^x}\right)^2\frac{J_1^z+J_2^z}{J_2}
\end{equation}
A non polarized XXZ chain (corresponding to commensurate filling for the bosons)
is in a LL phase for $|J^z|/J<1$ and has a gap otherwise. For the
pseudo-spin Hamiltonian this means that as we increase the
staggered field the model becomes gapless for a critical value of
order $h_x\propto J^z_\perp$.
One recovers the BKT transition with an effective
universal critical LL exponent $\tilde{K}_c=1/2$ and a dynamical exponent
$z=1$. Using (\ref{eq:mapping})
and $S_\beta^+\propto\psi_1^\dagger\psi_2^\dagger$ for the bosons
it is easy to check from (\ref{eq:correl}) that the original bosonic ladder has
a universal LL parameter $K_s^*=1$.
The diagonalization on
each rung shows that $S^x_{\alpha,i}$ has a non-zero mean value,
which we interpret as an order in the field $\theta_\alpha$.
These results for the transition are
coherent with the previous large transverse hopping analysis, which
gives some confidence that we indeed capture the correct critical behavior.
The two above analysis extend easily to the situation where the chemical potential
does not ensure particle hole symmetry. In that case, this leads to either
a magnetic field in the spin representation or an extra chemical potential
in the effective single bosonic chain limit. In each case the BKT transition
becomes a commensurate-incommensurate one \cite{schulz_cic2d},
with a universal LL parameter
$K_s^c=1/2$ and a dynamical exponent $z=2$.
The other properties at the transition (compressibility, etc.) both
for the commensurate and incommensurate case
can be obtained in a similar way than for the simple Mott transition for
bosons \cite{giamarchi_mott_shortrev}.
A summary of the critical behavior is shown on Fig.~\ref{fig:phasediag}.

Various physical observables can be computed. In the superfluid
phase the symmetric modes are described by a LL with a parameter
$K_s$, whereas the $\theta_a$ mode is gapped. Thus most
correlation functions involving the superfluid phase decay as power law.
For example
\begin{eqnarray} \label{eq:correl}
\langle\psi_\alpha^\dagger(r)\psi_\alpha(0)\rangle
&\propto& \left(\frac1r\right)^{\frac1{4K_s}} \\
\langle\psi_1^\dagger\psi_2^\dagger(r)\psi_1\psi_2(0)\rangle
&\propto& \left(\frac1r\right)^{\frac1{K_s}}
\end{eqnarray}
In the Mott phase $\phi_s$ and $\phi_a$
are ordered so the correlations functions such as (\ref{eq:correl})
decay exponentially. One of the most important difference between the
two phases are of course the transport properties. The Drude weight
is zero in the Mott insulating phase, whereas it is
given by ${\cal D} = 2 u_s K_s$ in the superfluid one, with a discontinuous
jump at the transition.  Far from the transition
one can obtain $K_s$ from the RG. We use a two scale renormalization
and cut the renormalization due to the umklapp when the
the transverse hopping reaches a value of order one. This gives
\begin{eqnarray}
{\cal D} = 2 u_s K_s \left(1 - \frac{C}{K_s}
\left[\tilde{t}_{\perp}^{\frac{-4K(2-K)}{4K-1}}-1\right]\right)
\end{eqnarray}
where $C$ is a constant of order unity. Thus the Drude weight
decreases as one approaches the transition by reducing $t_\perp$.

The temperature and frequency dependence of the conductivity can
also be extracted from the RG. The high
frequency behavior may be perturbativily computed from the umklapp
scattering operator $g_u(\cos{2\phi_1}+\cos{2\phi_2})$. This leads
to $\sigma(\omega)\propto g_u^2 \omega^{2K-5}$, similar to
the result for a single chain \cite{giamarchi_mott_shortrev}.
However for frequencies less than the gap in the
antisymmetric mode, $\Delta_a$, the original umklapp operator has been
renormalized nearly to zero. The dominant scattering operator is
$\cos(\sqrt8\phi_s)$ leading to $\sigma(\omega) \propto g_u^4 \omega^{4K_s-5}$.
In the Mott phase there is a gap $\Delta_s$ in the symmetric mode
below which these behaviors are cut, whereas it extends to $\omega=0$ in
the superfluid phase which has in addition the Drude weight at zero frequency.
\begin{figure}
\includegraphics[width=9cm]{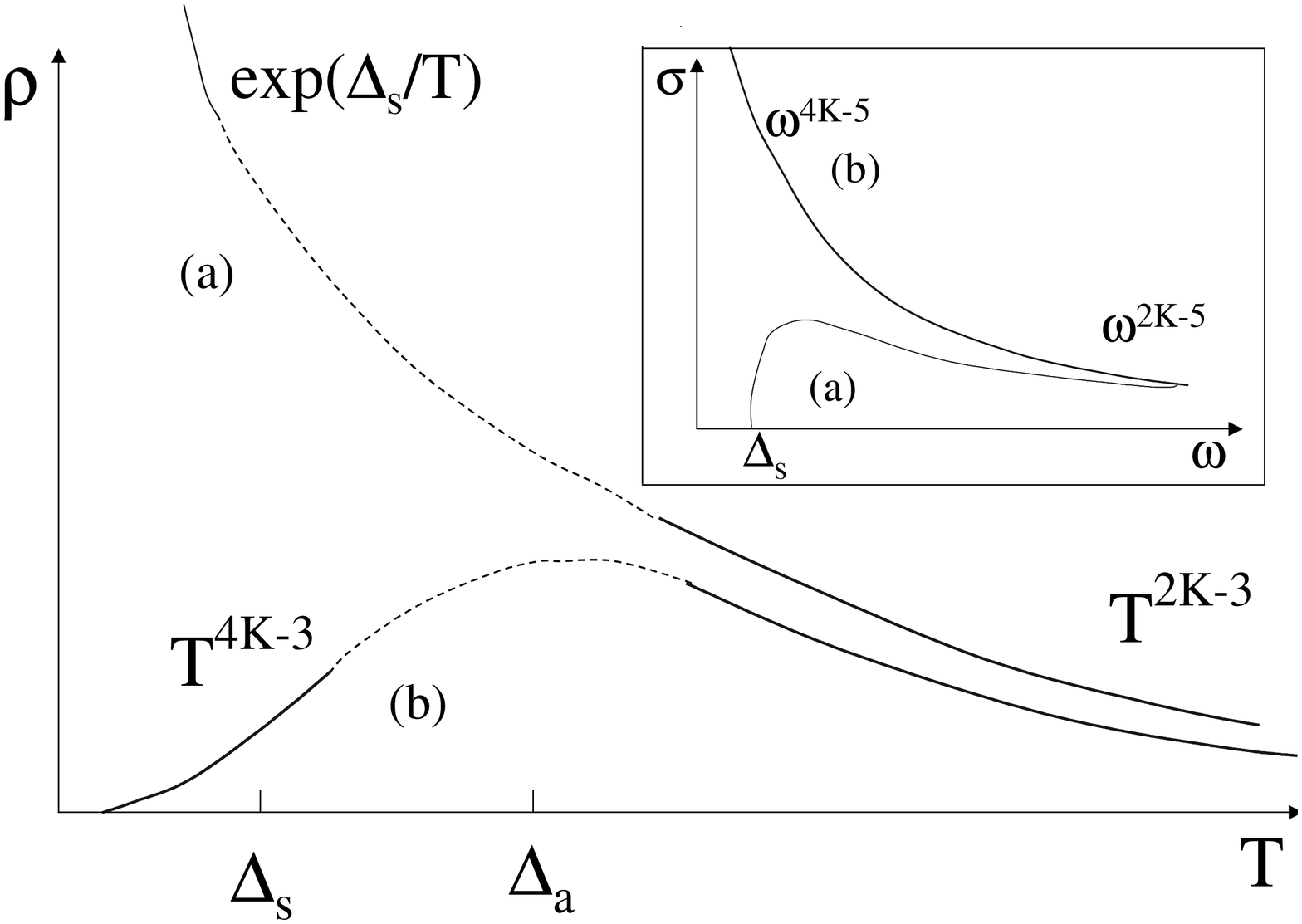}
\caption{\label{fig:rholadder} The curve $(a)$ is a schematic vue
of the resistivity for the ladder in the mott-insulating state,
with an activated behavior for temperatures lower than the charge
gap $\Delta_s$. The curve $(b)$ is for a superfluid ladder (shown for $K<1.5$) (see
text). The dashed part of the curves represents the cross-over
region between low and high temperature. The insert shows the
optical conductivity (shown for $K<1.25$).}
\end{figure}
Similar behavior is obtained for the temperature dependence of the
conductivity \cite{giamarchi_mott_shortrev}:
\begin{eqnarray} \label{eq:tempcond}
\rho(T) &\propto&  T^{2K-3} \qquad,\qquad T > \Delta_a \\
\rho(T) &\propto&  T^{4K-3} \qquad,\qquad T < \Delta_a
\end{eqnarray}
In the Mott phase one recovers the familiar activated exponential behavior
when $T < \Delta_s$. Both the dc and ac conductivity are shown on Fig.~\ref{fig:rholadder}.
Note that for $K_s>1.5$ the temperature dependence is non monotonous in
the superfluid phase, whereas in the insulating one the resistivity
would start to increase with decreasing temperature even well above the
Mott gap.

The above predictions could be checked either in numerical simulations
or in experimental systems such as the JJ arrays. In numerical simulations
the LL parameters could be extracted in a way similar to the one that was used
for the single bosonic chain \cite{kuhner_bosehubbard}. The gaps, superfluid correlation functions and
the Drude weight are probably the most easily checkable quantities.
For JJ arrays, as for the single chain the transition we predict should
be visible as a metal-insulator transition in a transport experiment.
One possible way to control $t_\perp$ could be to use a magnetic field
\cite{orignac_vortex_ladder}.

\begin{acknowledgments}
We would like to thank T. Kuehner, H. Monien, E. Orignac and G.T. Zimanyi
for useful discussions. This work has been supported in part by NATO grant 971615.
\end{acknowledgments}


\end{document}